\documentstyle[12pt]{article}
\begin{document}
\begin{titlepage}
\title{Leptons, quarks, and their antiparticles:\\ a phase-space view}
\author{
{Piotr \.Zenczykowski 
}\\
{\em Division of Theoretical Physics},
{\em Institute of Nuclear Physics},\\
{\em Polish Academy of Sciences},\\
{\em Radzikowskiego 152,
31-342 Krak\'ow, Poland}\\
e-mail: piotr.zenczykowski@ifj.edu.pl
}
\maketitle
\begin{abstract}
Recently, a correspondence has been shown to exist between
the structure of a single Standard Model generation 
of elementary particles and the 
properties of the Clifford algebra of nonrelativistic phase space.
Here,
 this correspondence is spelled out
in terms of phase-space variables.
Thus,
a phase-space  
interpretation of the connections between leptons, quarks and their
antiparticles is proposed, in particular providing a timeless alternative to 
the standard St\"uckelberg-Feynman interpretation.
The issue of the additivity of canonical momenta is 
raised and argued to be intimately related to the unobservability of free
quarks and the emergence of mesons and baryons.\\

Keywords:\\ Quantum geometry, phase space, Clifford algebra, structure of
Standard Model generation

\end{abstract}
\vfill

PACS numbers: 11.30.-j; 04.60.Pp; 12.60.-i

\end{titlepage}

\section{Introduction}

In the standard formalism used for the description of elementary particles, 
transformations between  observed particles are given within 
the context of various symmetry groups, such as $SU(2)$, $SU(3)$, etc.
In particular, transformations
between particles and antiparticles are effected via
complex conjugation, with antiparticles belonging to complex-conjugate
representations of the symmetry groups in question.
Complex conjugation is also needed when discussing time reversal, e.g. in the
time-dependent Schr\"odinger equation.
Thus, the existence of particles and antiparticles 
appears to be closely related to the
existence of the macroscopic concept of external time.
And indeed, St\"uckelberg and Feynman interpreted antiparticles as `particles
moving backward in time'.

In quantum gravity, however,
 the concept of external time is 
$-$ as argued by many
(e.g. \cite{Wheeler,Rovelli1991,Kiefer}) $-$  presumably
the first notion to be discarded.  In fact, `time' is never 
measured directly, we always measure {\it positions}, 
be it positions of clock hands, or positions of other groups of particles.
Ultimately, we deal with relative positions 
 only. 
Consequently, the `flow of time' 
may be replaced by the observed  regular changes in the
correlations between positions of  (groups of) particles \cite{Barbour1}
 (regardless of which of our senses they affect). 
 This is what astronomers actually do: 
they define time through 
changes in the positions of astronomical objects \cite{BarbourNatureoftime}.
Obviously,  at the deep level and
irrespectively of this procedure,
there should exist something that
time originates from.
If it is not a classical time, it
might be its quantized version, a quantized version of changes in 
objects' relative positions, or something much more involved.

Now, when the external time of the Schr\"odinger equation vanishes from the formalism
(in some approximation being presumably superseded by such 
changes in particles' positions),
 the rationale for complex numbers in quantum gravity seems to vanish as well \cite{Barbour}. 
What happens then to the particle-antiparticle quantum degree of freedom?
Obviously, we do not expect it to vanish. 
Quantum gravity, just as the ordinary quantum description of Nature, 
is believed to be complex. In fact, in any quantum approach 
the `$i$'  should enter the game
through the quantum-mechanical position-momentum commutation relations,
with position and momentum treated very symmetrically.
The continuing presence of `$i$' in the mathematical structure not involving
explicit time variable suggests that it should  be possible to interpret
the existence of both particles and antiparticles also
from the point of view of phase space alone, i.e.
without an introduction of the concept of (external) time. 
Such an interpretation would shift our vantage point 
significantly, and 
could be useful 
in our search for a proper approach to (timeless) quantum gravity.

Keeping in mind that
quantum mechanics `lives' in phase space,
I argued in a couple of recent papers 
(e.g. \cite{APPB2007},\cite{Zen2008},\cite{JPA}) 
that viewing phase space as an arena of physics
may provide a clue to `space quantization'.
At the level of basic assumptions, the relevant ideas 
differ  from those 
of the formalism standardly known under the name
of `phase-space quantization' (see e.g. \cite{Zachos}), however.
Specifically, papers \cite{APPB2007},\cite{Zen2008},\cite{JPA} consider
a Dirac-like linearization of 
${\bf x}^2+{\bf p}^2$, with position and momentum satisfying standard
commutation relations.
This leads to the apperance of a previously undiscussed
discrete structure: the Clifford algebra
of nonrelativistic phase space endowed with form ${\bf x}^2+{\bf p}^2$. 
This algebra is interesting because a part of it might be
identified with the charge (isospin, hypercharge) structure of a single generation of elementary fermions in
the Standard Model \cite{Harari}. In fact, the
transition from nonrelativistic quantum
phase space to its Clifford algebra may be regarded 
as a small step towards the quantization of `space'. 
The main difference with respect to other approaches
to quantum gravity is that here one deals
with the phase space, not with the ordinary configuration space.
In such a scheme, the spatial and internal quantum numbers of elementary particles are thought of as revealing the quantum layer underlying the geometry of phase space \cite{JPA},
considered as an arena of physics. 
This line of thinking leads to the idea
that understanding the whole quantum structure 
manifested via the spectrum of elementary particles (in particular, their
quantized masses and other parameters) should be 
 essentially equivalent to understanding quantum gravity.
 An argument that the effects of quantum gravity should appear only when  
 the Planck scale is reached
 is not really valid. 
 As stressed by Kiefer \cite{Kiefer}, quantum gravity effects
 `are not restricted to this scale a priori. The superposition principle
 allows the formation of non-trivial gravitional quantum states at any scale.
 (...) there may be situations where the quantum nature of
 gravity is visible $-$ even far away from the Planck scale'.
  Kiefer then points out that 
   such effects may be seen 
  at cosmological scales, far beyond the scale of elementary particles. 
  In fact, as long as we are dealing with quantum description, the issue of scale is
  simply irrelevant: 
  quantum physics suggests 
that the familiar classical concepts of space and
time are emergent concepts only, nonexisting  (in the classical form)
at the underlying
quantum level.
Aside from the above arguments, a more formal link to gravity 
seems to be present in the phase-space approach as well: 
when viewed from our 3D vantage point,
 the odd part of the Clifford algebra in question 
 contains, besides scalars and vectors, also {symmetric} $SO(3)$ tensors 
 of rank 2 \cite{JPA}.

In \cite{APPB2007},\cite{Zen2008} various
transformations between the lepton and quark sectors of the Clifford algebra
were considered. 
In this note, I would like to discuss  the meaning of these transformations
and of their phase-space counterparts somewhat further,
so that an
understanding of the whole approach
in terms of the concepts of positions and momenta may be deepened,
including an interpretation of the
relation between leptons, quarks, and their antiparticles 
in phase-space
terms.

While our phase-space approach originates from the wish to treat
momenta and positions in a more symmetric way \cite{Born1949}, it is obvious that
one cannot restore full symmetry between them: the momenta and positions would be
then completely alike, which would be in conflict with the
observed physical reality.
In fact, putting aside the connection between the standard concept of mass 
and momentum (see \cite{JPA}), an important difference between  momenta and positions 
can be identified when one considers composite systems of ordinary particles
(by ordinary particles I mean those particles which may be {\it individually}
observed,
like leptons, hadrons, 
and systems built thereof, but not quarks).
For such systems,
one observes that
the physical momentum of the system is obtained by simply {\it adding}
the physical momenta of its components, independently of where 
(in the classical case) the
particles are. 
On the other hand, an (approximate) description of the physical
position of the whole system is best provided by certain {\it average}
of individual positions of its components (i.e., by their center of mass).
Keeping in mind that one of our goals is to
describe hadrons as composite systems made of quarks,
it is natural to ensure that additivity of the just discussed or similar type be 
included in the description involving quarks.
Towards the end of this paper we shall show that the r\^{o}le played by the
additivity principle
seems to be an important one: if the proposed ideas are basically correct, 
it is intimately related to quark unobservability and the emergence of mesons
and baryons.

\section{Nonrelativistic phase space and its Clifford algebra}
\subsection{Phase space} 
Our approach is driven by
the old wish to reduce dynamics to geometry, which
 motivates the search for a `pregeometric algebra',
a quantum-level counterpart of geometry.
As already mentioned, this quantum-level algebra 
should be capable of relating to geometry
such quantum features of Nature as the existence
of internal quantum numbers and the quantization of particle masses.
Thus, the corresponding effective quantum description in phase space
should involve, besides the Planck constant, also
a dimensional constant setting the mass scale.
These constants together (plus the speed of light) determine
 an absolute scale of momenta and distances, and
 permit measuring the distances in units of momenta and vice versa. 

For the above reasons, it was argued in \cite{APPB2007} that a natural generalization of the $O(3)$ symmetry of
ordinary 3D space to the case of nonrelativistic quantum phase space 
consists in considering the invariance of
\begin{equation}
\label{Inv1a}
{\bf x}^2+{\bf p}^2, 
\end{equation}
(originally identified with Hamiltonian, but now viewed in geometrical terms),
with ${\bf x}$ and ${\bf p}$ being the operators of {physical} positions and momenta,
subject to the condition that 
\begin{equation}
\label{Inv1b}
[x_k,p_l]=i\delta_{kl}
\end{equation} is also
invariant.
With the absolute scales of positions and momenta fixed, 
the symplectic
rescalings admitted by Eq. (\ref{Inv1b}) are obviously no longer allowed.

As is well known, the resulting symmetry is 
then $U(1) \otimes SU(3)$. Note that 
this symmetry is in fact present already 
in the strictly classical case, with
Poisson brackets in place of the commutators.

A straightforward generalization of the set of 
Eqs (\ref{Inv1a},\ref{Inv1b}) is obtained by realizing that, since $i$ is defined only up to a sign,
Eq (\ref{Inv1b}) may be replaced by
\begin{equation}
\label{Inv1c}
[x_k,p_l]=-i\delta_{kl},
\end{equation}
leading to a different copy of the 
$U(1) \otimes SU(3)$ symmetry.

It should also be stressed that the operation of complex conjugation does not connect
Eq. (\ref{Inv1b}) with, Eq. (\ref{Inv1c}).
Indeed, using the standard representation
$p_k=-i\frac{d}{dx_k}$ one obtains that complex conjugation
corresponds to
\begin{eqnarray}
\label{xpicc1}
i\to -i,~~~ &{\bf x}\to {\bf x}, & ~~~{\bf p} \to -{\bf p}.
\end{eqnarray}
Under complex conjugation the set of  Eqs (\ref{Inv1a},\ref{Inv1b}) (or
alternatively, the set of  Eqs (\ref{Inv1a},\ref{Inv1c}))
 remains therefore invariant.

\subsection{Clifford algebra and charges of elementary fermions}
Linearization of expression (\ref{Inv1a}) \`a la Dirac (with commuting positions
and momenta) leads to Clifford algebra
built from the basic elements $A_k$ and $B_l$, associated with momentum $p_k$ and
position $x_l$ respectively. We use
the following explicit representation:
\begin{eqnarray}
A_k&=&\sigma_k\otimes \sigma_0 \otimes \sigma_1, \nonumber\\
B_k&=&\sigma_0\otimes \sigma_k \otimes \sigma_2, \nonumber\\
\label{AkBkB7}
B_7\equiv iA_1A_2A_3B_1B_2B_3&=&\sigma_0\otimes \sigma_0 \otimes \sigma_3,
\end{eqnarray}
where $B_7$ is the 7th anticommuting element of the algebra.
\footnote{
The product $\Pi_{k=1}^3A_kB_k$ is the 6D generalization of 
$A_1B_1$ relevant for
the 2D phase space. For the latter case, an identification of such a product 
with the imaginary unit 
was proposed in \cite{Pavsicbook}.}

In \cite{APPB2007} it was shown that when the
position-momentum commutation relations are accepted, one obtains
\begin{equation}
R^{tot}\equiv({\bf A}\cdot {\bf p}+{\bf B}\cdot {\bf x})
({\bf A}\cdot {\bf p}+{\bf B}\cdot {\bf x})
=({\bf p}^2+{\bf x}^2)-\frac{i}{2}[A_k,B_k],
\end{equation}
(suumation over repeated indices implied)
where the last term comes about because $x_k$ and $p_k$ do not commute.
In \cite{APPB2007} it was then furthermore proposed that electric charge $Q$ is equal to 
 operator $\frac{1}{6}R^{tot}B_7$, evaluated for the lowest level
of ${\bf p}^2+{\bf x}^2$, i.e.
\begin{equation}
\label{QGMN}
Q= \frac{1}{6}\left(({\bf p}^2+{\bf x}^2)_{lowest}-\frac{i}{2}[A_k,B_k]\right)
B_7\equiv I_3+\frac{Y}{2},
\end{equation}
with  (weak) isospin $I_3$ and (weak) hypercharge $Y$ given by
\begin{eqnarray}
I_3&=&\frac{B_7}{2}\nonumber\\
\label{I3andY}
Y&=&-\frac{i}{6}[A_k,B_k]B_7=
\frac{1}{3}\sigma_k\otimes\sigma_k\otimes\sigma_0\equiv \sum_k Y_k.
\end{eqnarray}
The eigenvalues of $I_3$ are $\pm 1/2$, those of `partial hypercharges' $Y_k$ 
are $\pm 1/3$, while for the hypercharge $Y$ we have $-1,+1/3,+1/3,+1/3$, corresponding to a lepton and a
triplet of quarks respectively.
Eq. (\ref{QGMN}) yields then the charges of all eight particles of a
single generation of the Standard Model.

\section{Isospin reversal and charge conjugation}
Transformations in the phase space and in its Clifford algebra
 are related by the
requirement of the invariance of ${\bf A}\cdot {\bf p}+{\bf B}\cdot{\bf x}$.
Below we shall use this invariance to provide phase-space interpretations of the
connections between the particles of a single Standard Model generation.

\subsection{Isospin reversal}
Consider the following reflection in the algebraic counterpart of phase space:
\begin{eqnarray}
\label{phspreflection}
{\bf A'}=I_1{\bf A}I_1^{-1}={\bf A},&
~~~&{\bf B'}=I_1{\bf B}I_1^{-1}=-{\bf B},
\end{eqnarray}
where $I_n=\sigma_0\otimes\sigma_0\otimes\sigma_n$.
Under the above operation,
we have $B_7 \to -B_7$, $i \to i$, and, consequently,
this transformation swaps $I_3=\pm \frac{1}{2}$ sectors, while keeping
hypercharge unchanged:
\begin{eqnarray}
\label{phspreflect}
I'_3=-I_3, &~~~ & Y'=Y.
\end{eqnarray}
Upon requiring the invariance of the expression
${\bf A} \cdot {\bf p}+{\bf B} \cdot {\bf x}$
one finds that Eq. (\ref{phspreflection}) 
corresponds to the following transformation in phase space:
\begin{eqnarray}
\label{phsprefl}
{\bf p} \to {\bf p'}={\bf p},  &
~~~~~{\bf x} \to {\bf x'}= -{\bf x}. &
\end{eqnarray}
Alternatively, one might consider
 ${\bf A'}=I_2{\bf A}I_2^{-1}=-{\bf A}$, ${\bf B'}=I_2{\bf B}I_2^{-1}={\bf B}$, and 
 ${\bf p'}=-{\bf p} $, ${\bf x'}= {\bf x}$, 
  which is related to 
(\ref{phspreflection}) via ordinary 3D reflection.
Then, Eq. (\ref{phspreflect}) still holds.
Both types of reflection correspond to the transformation
\begin{eqnarray}
[x_k,p_l]=i\,\delta_{kl} & \to & [x_k,p_l]=-i\,\delta_{kl},
\end{eqnarray}
with ${\bf p}^2+{\bf x}^2$ unchanged, i.e. we recover the other copy of $U(1) \otimes SU(3)$ (cf Eq. (\ref{Inv1c})).
We prefer to work with phase-space reflection defined in Eq. (\ref{phspreflection}) 
as it does not affect 
the momentum of the particle under consideration.

If lepton $L$ of isospin $I_3=+1/2$ corresponds 
to the following division of the six basic elements of
Clifford algebra into the counterparts of 
canonical 
momenta, and counterparts of canonical positions: 
\begin{equation}
\label{leptonplusconv}
({\bf A}^L,{\bf B}^L) =({\bf A}, {\bf B}),
\end{equation}
with (the operators of) canonical momenta ${\bf p}^L$ (positions ${\bf x}^L$) 
identified with (the operators of) physical momenta ${\bf p}$
(positions ${\bf x}$), i.e.:
\begin{equation}
\label{pxplus}
({\bf p}^L,{\bf x}^L)=({\bf p},{\bf x}),
\end{equation}
then its partner $L'$ of isospin $I_3=-1/2$ corresponds to
\begin{equation}
\label{leptonminusconv}
({\bf A}^{L'},{\bf B}^{L'}) =({\bf A}, -{\bf B}),
\end{equation}
and
\begin{equation}
\label{pxminus}
({\bf p}^{L'},{\bf x}^{L'})=({\bf p},-{\bf x}).
\end{equation}
The choice of Eq (\ref{phspreflection}) for the representation of isospin
reversal  ensures that the addition
of the canonical momenta of leptons is equivalent to the
addition of their physical momenta, independently of their eigenvalues of $I_3$.

\subsection{{\label{chargeconj}}Charge conjugation}
As is well known, charge conjugation is closely related to complex conjugation.
A possible corresponding operation in phase space was given 
in Eq. (\ref{xpicc1}).
As in the case of phase-space reflection
we prefer to keep the momentum unchanged 
(so that we go from a particle of a given
physical momentum to its antiparticle of the {same} physical momentum)
and {\it define}  the 
operation of charge conjugation in phase space via:
\begin{eqnarray}
\label{xpicc2}
{\bf \bar{p}}={\bf p}, &~~~~{\bf \bar{x}}= -{\bf x}, & ~~~~
\bar{i}=i^*=-i,
\end{eqnarray}
as is also obtained from the straightforward 
application of complex conjugation if one takes $x_k=i\frac{d}{dp_k}$, with $p_k$ real.
As before, the reason for this choice of the definition of charge conjugation is
that the addition of physical momenta of particles and antiparticles is now
straightforward: one does not have to employ any specific method
 of inverting the signs of antiparticle momentum 
variables. In Clifford algebra
the operation of charge conjugation should be obtained 
 from $i \to i^*=-i$ through:
\begin{eqnarray}
\label{ChconjClAlg}
{\bf A} \to {\bf A}^*, &~~~~~&{\bf B} \to {\bf B}^*.
\end{eqnarray}
Then, we get
\begin{eqnarray}
{\bf A} \cdot {\bf p} +{\bf B} \cdot {\bf x} & \to &
{\bf A^*} \cdot {\bf p} - {\bf B^*}\cdot {\bf x}.
\end{eqnarray} 
The condition of the invariance of 
${\bf A} \cdot {\bf p} +{\bf B} \cdot {\bf x}$ under charge conjugation
may be satisfied if it is possible to find a unitary transformation such that
${\bf A}^* \to {\bf A},  {\bf B}^* \to -{\bf B}$,
i.e.  such ${C}$ that:
\begin{eqnarray}
\label{properptoap}
{\bf \bar{A}}&=&{C}{\bf A}^*{C}^{-1}={\bf A},\nonumber\\
{\bf \bar{B}}&=&{C}{\bf B}^*{C}^{-1}=-{\bf B},
\end{eqnarray}
so that a counterpart of Eq. (\ref{xpicc2}) is obtained:
\begin{eqnarray}
\label{ChconjAB}
{\bf \bar{A}}={\bf A}, &~~~~~{\bf \bar{B}}=-{\bf B},
&~~~~~({\rm and}~~~\bar{i}= i^*=-i).
\end{eqnarray}
One may check that 
\begin{equation}
\label{chconjdef}
{C}={C}^{-1}=\sigma_2 \otimes \sigma_2 \otimes \sigma_3.
\end{equation}
possesses the required property.
\footnote
{In \cite{APPB2007}
the operation of charge conjugation was defined in a similar way, but with
a different ${C}$, 
proportional to $\sigma_2 \otimes \sigma_2 \otimes \sigma_2$.
I now find the latter expression inadequate for a
couple of reasons, the requirement of the invariance of ${\bf A} \cdot {\bf p}+
{\bf B} \cdot {\bf x}$ among them.
Adopting prescription (\ref{chconjdef}) changes the sign with which ${\bf B}$
(and $B_7$) transforms under charge conjugation
(when compared to \cite{APPB2007}), but does not affect the conclusions of
\cite{Zen2008} since the eigenvalues of $B_7$ and $-B_7$ are
identical.}
From Eq. (\ref{AkBkB7}) one then finds that $B_7 \to B_7$, whence
\begin{eqnarray}
I_3 \to I_3, &~~~& Y \to Y. 
\end{eqnarray}
Thus, the exponent in $U=\exp(iQ)$ changes as follows
\begin{eqnarray}
iQ=i(I_3+Y/2)& \to & -i(I_3+Y/2)\stackrel{def}{=}i\bar{Q}=
i(\bar{I}_3+\bar{Y}/2),
\end{eqnarray}
i.e. the antiparticles have opposite charges, isospins and hypercharges:
\begin{eqnarray}
\bar{Q}=-Q,~~~&~~~\bar{I}_3=-I_3,~~~&~~~\bar{Y}=-Y.
\end{eqnarray}

The antiparticle to any lepton $L$ of isospin $I_3=+1/2$,
i.e. an antilepton $\overline{L}$ of isospin 
 $\bar{I}_3=-1/2$, corresponds to
\begin{equation}
\label{antileptonconv}
({\bf A}^{\bar{L}},{\bf B}^{\bar{L}},i^*) =({\bf A}, -{\bf B},-i),
\end{equation}
and
\begin{equation}
\label{antipx1}
({\bf p}^{\bar{L}},{\bf x}^{\bar{L}})=({\bf p},-{\bf x}),
\end{equation}
where ${\bf p}$ and ${\bf x}$ are the (operators of)
 physical momenta and positions of an
antiparticle.
Thus, {\it apart from} $i \to -i$, 
the difference between particles and antiparticles
amounts again to a reflection in position space.
We stress that the difference between the phase space reflection of 
Eq. (\ref{phspreflection},\ref{phsprefl}) and the
 representation (\ref{xpicc2},\ref{ChconjAB}) 
 of charge conjugation in phase space is tiny:
it is just the change of the sign in front of {\it free-standing} $i$,
without affecting ${\bf p}$ or ${\bf x}$.
 The $\bar{I}_3=+1/2$ isospin partner $\bar{L'}$ of antilepton $\bar{L}$ 
 (of Eqs (\ref{antileptonconv},\ref{antipx1}))
 corresponds to
 \begin{eqnarray}
 ({\bf A}^{\bar{L'}},{\bf B}^{\bar{L'}},i^*) &=&({\bf A},{\bf B},-i),\\
  ({\bf p}^{\bar{L'}},{\bf x}^{\bar{L'}})&=&({\bf p},{\bf x}).
 \end{eqnarray}
 With the above definitions,
the additivity of the canonical momenta of leptons and/or antileptons
  is equivalent to the additivity of their
 physical
 momenta, irrespectively of whether we are dealing with particles, 
 antiparticles, or both particles and antiparticles, 
 and irrespectively of their values of $I_3$.
 
 \section{Lepton-quark transformations}
 Apart from the just discussed 
 discrete transformations corresponding to isospin reversal 
 and charge conjugation, 
 one may consider lepton-quark transformations. 
As shown in \cite{APPB2007},\cite{Zen2008}, the transformations from the lepton
sector (of $Y=-1$) to
the three quark sectors 
(defined by the sets of the eigenvalues $\pm 1/3$ of $Y_k$'s such that $Y=\sum_kY_k=+1/3$)
correspond to specific cases of those phase-space
$SO(6)$ transformations which
go beyond $U(1) \otimes SU(3)$
(and the
standard 3D rotations in particular). 
Among these `genuine' $SO(6)$ transformations there are, in particular, 
the following (pairs of)
rotations:
\begin{eqnarray}
\tilde{p}_1=p_1\cos \phi +x_3 \sin \phi, &\phantom{xxxxx}&
\tilde{x}_1=x_1\cos \phi +p_3 \sin \phi,\nonumber\\
\label{Fplus2}
\tilde{x}_3=x_3\cos\phi-p_1\sin\phi,
&&\tilde{p}_3=p_3\cos\phi-x_1\sin\phi,
\end{eqnarray}
which leave $(x_2,p_2)$ unchanged: $\tilde{p}_2=p_2$, $\tilde{x}_2=x_2$.
Transformation between the lepton sector and the colour-$2$ quark
sector  
is obtained   when a specific condition is imposed 
on Eqs (\ref{Fplus2}). This condition amounts to the requirement 
that the position-momentum commutation
relations in new variables, i.e.
\begin{equation}
[\tilde{x}_k,\tilde{p}_l]=i\Delta_{kl}
\end{equation} 
be diagonal, 
and that new canonical positions $\tilde{x}_k$ 
(and new canonical momenta $\tilde{p}_k$) 
commute among themselves. As discussed in \cite{APPB2007},
a nontrivial case (corresponding to a quark of colour-$2$) 
is obtained  for $\phi = \pm \pi/2$,
which yields
\begin{equation}
\Delta=\left[
\begin{array}{ccc}
-1 & 0 & 0\\
0 &1&0\\
0&0&-1
\end{array}
\right],
\end{equation}
with similar formulas for the two analogs of Eqs (\ref{Fplus2}) corresponding to
(pairs of) rotations leaving $(x_3,p_3)$ and $(x_1,p_1)$ unchanged.
Eqs (\ref{Fplus2}) for $\phi=\pm\pi/2$ define what we mean by canonical momenta
and positions for quarks of colour $2$ (up to a redefinition corresponding to
$\phi = \pi/2 \leftrightarrow \phi= -\pi/2$, see later).

Putting aside the above three types of transformations 
(which interchange positions and momenta)
and the nine types of 
transformations corresponding to
$U(1) \otimes SU(3)$, one is left with the three remaining types of $SO(6)$
transformations. These are similar to ordinary 3D rotations, the difference
being that rotations in momentum and position spaces are now performed in opposite
senses:
\begin{eqnarray}
\tilde{p}_1=p_1\cos \phi -p_3 \sin \phi, &\phantom{xxxxx}&
\tilde{x}_1=x_1\cos \phi +x_3 \sin \phi,\nonumber\\
\label{Fminus2}
\tilde{p}_3=p_3\cos\phi+p_1\sin\phi,
&&\tilde{x}_3=x_3\cos\phi-x_1\sin\phi,
\end{eqnarray}
with $p_2,x_2$ unchanged.
The diagonality condition again requires $\phi =\pm \pi/2$.
In both cases, therefore, only specific dicrete cases of 
 the `genuine' phase-space rotations are allowed.\\

Phase-space transformations of Eqs (\ref{Fplus2},\ref{Fminus2}) have their obvious counterparts
in Clifford algebra. Thus, as discussed in \cite{APPB2007},
the set of all discrete transformations, 
leading from any algebra element $Z$ 
to its colour-$n$ quark counterpart $Z^{Qn}$,
consists of 
the following four alternatives 
(no summation over {underlined} indices):
\begin{eqnarray}
\label{QfromL1}
Z^{Qn}&=&\left\{
\begin{array}{c}
{\cal{R}}_{0{\underline{n}},\pm}
Z{\cal{R}}_{0{\underline{n}},\pm}^{-1},\\
\\
{\cal{R}}^{\dagger}_{0{\underline{n}},\pm}
Z({\cal{R}}^{\dagger}_{0{\underline{n}},\pm})^{-1},
\end{array}\right.
\end{eqnarray}
with
\begin{eqnarray}
{\cal{R}}_{0n,\pm}&=&\exp \,(i\phi F_{\pm n})|_{\phi=\pi/2}
=\exp \,(i\frac{\pi}{2}F_{\pm n})
=1+iF_{\pm n}-(F_{\pm n})^2,\nonumber\\
{\cal{R}}^{\dagger}_{0n,\pm}&=&\exp \,(i\phi F_{\pm n})|_{\phi=-\pi/2}
=\exp \,(-i\frac{\pi}{2}F_{\pm n})
=1-iF_{\pm n}-(F_{\pm n})^2,
\end{eqnarray}
generated by the `genuine'
$SO(6)$ / $SU(4)$ generators:
\begin{eqnarray}
F_{+n}&=&-\frac{i}{4}\epsilon_{nkl}\,[A_k,B_l]=
\frac{1}{2}\,\epsilon_{nkl}\,\sigma_k\otimes\sigma_l\otimes\sigma_3\nonumber\\
F_{-n}&=&-\frac{i}{4}\epsilon_{nkl}(B_kB_l-A_kA_l)=
\frac{1}{2}\,(\sigma_0\otimes \sigma_n-\sigma_n\otimes \sigma_0)
\otimes \sigma_0,
\end{eqnarray}
corresponding (for $n=2$) to (\ref{Fplus2}) and (\ref{Fminus2}) respectively,
and satisfying $(F_{\pm n})^3=F_{\pm n}$.
The four alternatives of Eq. (\ref{QfromL1}) exist because
operators $Y$ and $I_3$ are even in $A_k$, $B_l$, and therefore different
transformations of $A_k$, $B_l$ may lead to the same result for $Y$ and $I_3$.
In particular, for a given $n$, all four transformations (\ref{QfromL1})
 change the partial hypercharges $Y_k$ (and the isospin $I_3$) in exactly the
 same way, which amounts to
  interchanging the $Y=-1$ (lepton) sector with the
  $Y=+1/3$ (quark) colour-$n$ sector.
  (For example, for ${\cal{R}}_{0{{n}},+}$ and ${\cal{R}}^{(\dagger)}_{0{{n}},+}$
  this may be seen from Eqs (\ref{I3andY},\ref{eqAB},\ref{eqABdagger}).)

 When describing all three colours simultaneously, 
  one might in principle consider various combinations of 
  the ${\cal{R}}_{0{{n}},\pm}$- and 
 ${\cal{R}}^{\dagger}_{0{{n}},\pm}$-induced transformations,
  taking any of the four options in Eq. (\ref{QfromL1}) for a given colour
 (e.g. choosing either the set $\{{\cal{R}}_{0{{1}},+}$, ${\cal{R}}_{0{{2}},+}$, 
 ${\cal{R}}_{0{{3}},+}\}$, or the set $\{{\cal{R}}_{0{{1}},+}$, ${\cal{R}}_{0{{2}},-}$,
  ${\cal{R}}^{\dagger}_{0{{3}},+}\}$, or..., etc.). 
    Full symmetry between the three  
 directions 
 for both position and momentum, as
 observed in our 3D world, 
 requires however that we admit only such combinations
 which do not depend on our way of labelling
 the three directions with numbers $1$, $2$,
 $3$ (because the assignment of labels of, say, a right-handed system of
 coordinates to the three, yet unlabelled, directions of our ordinary 3D space
 is arbitrary).
 This restricts our study to four sets of transformations only,
 each set
 specified by the sense of the rotation by $\pi/2$ (i.e. $\phi = \pm \pi/2$), and
 by the type of $SO(6)$/$SU(4)$ generator used (either $F_{+n}$ or $F_{-n}$).
  
   Below
 we will consider the action of the transformations 
 of Eq. (\ref{QfromL1}) upon the algebraic counterparts
 of momenta and positions, i.e. upon
 elements $A_k$ and $B_l$. 
 We start with the set of 
 ${\cal{R}}_{0{{n}},+}$-induced
 transformations.

 \subsection{{\label{sectionRn+}}
 ${\cal{R}}_{0{{n}},+}$-induced transformations}

 The ${\cal{R}}_{0{{n}},+}$-induced ($\phi=+\pi/2$) 
 transformations 
 of $A_k$ and $B_l$ were evaluated in \cite{JPA} to be:
\begin{eqnarray}
{A}^{Qn}_{k}=
{\cal{R}}_{0\underline{n},+}A_k{\cal{R}}^{-1}_{0\underline{n},+}&=&\delta_{\underline{n}k}A_{\underline{n}}-
\epsilon_{nkm}B_m,\nonumber\\
\label{eqAB}
{B}^{Qn}_{k}={\cal{R}}_{0\underline{n},+}B_k{\cal{R}}^{-1}_{0\underline{n},+}&=
&\delta_{\underline{n}k}B_{\underline{n}}-
\epsilon_{nkm}A_m.
\end{eqnarray}
If we start with lepton $L$ of $I_3=+1/2$, i.e. with
 the counterparts of canonical momenta being
$({\bf A}^L,{\bf B}^L)=({\bf A},{\bf B})$, the above formulas yield
their coloured quark counterparts 
$({\bf A}^{Qn},{\bf B}^{Qn})=$
$(({A}^{Qn}_1,{A}^{Qn}_2,{A}^{Qn}_3),({B}^{Qn}_1,{B}^{Qn}_2,{B}^{Qn}_3))$.
The relevant expressions may be written in
 a transparent matrix form:
\begin{eqnarray}
A^Q\equiv 
\left[
\begin{array}{c}
{\bf A}^{Q1}\\
{\bf A}^{Q2}\\
{\bf A}^{Q3}
\end{array}
\right]
&=&
\left[
\begin{array}{ccc}
A_1&-B_3&+B_2\\
+B_3&A_2&-B_1\\
-B_2&+B_1&A_3
\end{array}
\right],\nonumber\\
&&\nonumber\\
\label{AgenerSU4R}
B^Q\equiv 
\left[
\begin{array}{c}
{\bf B}^{Q1}\\
{\bf B}^{Q2}\\
{\bf B}^{Q3}
\end{array}
\right]
&=&
\left[
\begin{array}{ccc}
B_1&-A_3&+A_2\\
+A_3&B_2&-A_1\\
-A_2&+A_1&B_3
\end{array}
\right],
\end{eqnarray}
 with Clifford algebra counterparts ${\bf A}^{Qn}$ for canonical momenta 
 and ${\bf B}^{Qn}$ for canonical positions gathered in row
 $n$ for the sector of colour $n$. 
The imaginary unit $i$ is of course unchanged
as ${\cal{R}}_{0\underline{n},+}i{\cal{R}}^{-1}_{0\underline{n},+}=i$ for any
$n$.
Since $i$ is unaffected, the pair of matrices ($A^Q$, $B^Q$) must
correspond to quarks. 
Furthermore, since 
${\cal{R}}_{0\underline{n},+}I_3{\cal{R}}^{-1}_{0\underline{n},+}=I_3$, again
for any $n$,
our quarks still have $I_3=+1/2$.

\subsection{${\cal{R}}^{\dagger}_{0{{n}},+}$-induced transformations}
The ${\cal{R}}^{\dagger}_{0{{n}},+}$-induced ($\phi=-\pi/2$) transformations 
 of $A_k$ and $B_l$ were found in \cite{JPA} to be:
 \begin{eqnarray}
{A}^{Qn(\dagger)}_{k}=
{\cal{R}}^{\dagger}_{0\underline{n},+}
A_k({\cal{R}}^{\dagger}_{0\underline{n},+})^{-1}&=&\delta_{\underline{n}k}A_{\underline{n}}+
\epsilon_{nkm}B_m,\nonumber\\
\label{eqABdagger}
{B}^{Qn(\dagger)}_{k}=
{\cal{R}}^{\dagger}_{0\underline{n},+}
B_k({\cal{R}}^{\dagger}_{0\underline{n},+})^{-1}&=&\delta_{\underline{n}k}B_{\underline{n}}+
\epsilon_{nkm}A_m.
\end{eqnarray}
If we start again with lepton $L$ of isospin $I_3=+1/2$, i.e. with
the counterparts of canonical momenta being
$({\bf A}^L,{\bf B}^L)=({\bf A},{\bf B})$, the above formulas yield
their coloured quark counterparts 
$({\bf A}^{Qn(\dagger)},{\bf B}^{Qn(\dagger)})$.
The corresponding matrices of the Clifford algebra counterparts
of canonical momenta and canonical positions are then:
\begin{eqnarray}
A^{Q(\dagger)}\equiv
\left[
\begin{array}{c}
{\bf A}^{{Q}1(\dagger)}\\
{\bf A}^{{Q}2(\dagger)}\\
{\bf A}^{{Q}3(\dagger)}
\end{array}
\right]
&=&
\left[
\begin{array}{ccc}
A_1&+B_3&-B_2\\
-B_3&A_2&+B_1\\
+B_2&-B_1&A_3
\end{array}
\right],\nonumber\\
&&\nonumber\\
\label{AgenerSU4Rdag}
B^{Q(\dagger)}\equiv
\left[
\begin{array}{c}
{\bf B}^{{Q}1(\dagger)}\\
{\bf B}^{{Q}2(\dagger)}\\
{\bf B}^{{Q}3(\dagger)}
\end{array}
\right]
&=&
\left[
\begin{array}{ccc}
B_1&+A_3&-A_2\\
-A_3&B_2&+A_1\\
+A_2&-A_1&B_3
\end{array}
\right].
\end{eqnarray}
Since, as in the previous case,
$i$ and $I_3$ are unaffected, 
the pair of matrices ($A^{Q(\dagger)}$, $B^{Q(\dagger)}$) 
$-$ just as the pair ($A^Q,B^Q$) $-$
must 
correspond to quarks of $I_3=+1/2$.

As discussed in \cite{APPB2007},\cite{Zen2008},\cite{JPA} and seen above, 
the $F_{+n}$-induced transformations lead to
 some of the
$A_k$'s being replaced by some of the $B_l$'s,
while their phase-space counterparts, i.e. the specific
cases of transformations (\ref{Fplus2}) 
(or their analogs)
lead to
some of physical
position variables 
playing the r\^ole of canonical momenta.

\subsection{${\cal{R}}_{0{{n}},-}$- and
${\cal{R}}^{\dagger}_{0{{n}},-}$-induced transformations}
As discussed in \cite{JPA}, for any fixed $n$ the 
${\cal{R}}_{0n,-}$-induced
transformations are related to the ${\cal{R}}_{0n,+}$ 
 via specific 
$U(1) \otimes SU(3)$ transformations
(with similar statement holding also for 
${\cal{R}}^{\dagger}_{0n,-}$ and ${\cal{R}}_{0n,+}^{\dagger}$).
The ${\cal{R}}_{0n,-}$-
(and ${\cal{R}}^{\dagger}_{0n,-}$-) induced transformations do not swap some $A_k$'s with some of the $B_l$'s, but
just rotate ${\bf A}$'s and ${\bf B}$'s in the opposite senses.
When the ordinary 3D rotations are also allowed, 
the effect of
${\cal{R}}_{0n,-}$- and
${\cal{R}}^{\dagger}_{0n,-}$-induced transformations
 is equivalent to admitting appropriate
transformations in the space of  canonical positions only.
Thus, they do not really bring in more symmetry between the momentum and position
coordinates that Max Born wanted so badly \cite{Born1949}.
Consequently, they are of no interest to us here.\\

\section{Representations}
For a fixed $n$ (take $n=1$) we have:
\begin{eqnarray}
{\bf A}^{Q1}=(A_1,-B_3,+B_2),~~\,&\phantom{xxx}&{\bf B}^{Q1}=(B_1,-A_3,+A_2),\nonumber\\
\label{quarkconventionAB}
{\bf A}^{Q1(\dagger)}=(A_1,+B_3,-B_2),&&{\bf B}^{Q1(\dagger)}=(B_1,+A_3,-A_2).
\end{eqnarray}
The above two possibilities for the counterparts of 
canonical momenta and positions for quark of colour $1$
are related by an ordinary 3D rotation by $\pi$ around the $1$st axis
in both ${\bf A}$ and ${\bf B}$ spaces
(alternatively, the connection  may be provided by appropriate reflections:
${\bf B} \to - {\bf B}$ for $A^Q$, and ${\bf A} \to -{\bf A}$ for $B^Q$).

The corresponding canonical momenta and positions are
\begin{eqnarray}
\label{quarkconventionpx}
{\bf p}^{Q1}=(p_1,-x_3,+x_2),~~\,&\phantom{xxx}&{\bf x}^{Q1}=(x_1,-p_3,+p_2),\nonumber\\
{\bf p}^{Q1(\dagger)}=(p_1,+x_3,-x_2),&&{\bf x}^{Q1(\dagger)}=(x_1,+p_3,-p_2).
\end{eqnarray}
As mentioned before, as long as $i$ is unchanged, both $({\bf A}^{Q1},{\bf B}^{Q1})$ and 
$({\bf A}^{Q1(\dagger)},{\bf B}^{Q1(\dagger)})$ represent algebraic counterparts of
canonical variables for quark of colour $1$ and isospin $I_3=+1/2$.
We now choose the first one of the two forms above
(i.e. the first rows in Eqs (\ref{quarkconventionAB},\ref{quarkconventionpx})) 
to represent quark of isospin $I_3=+1/2$ and colour $1$.
For arbitrary colour $n$, we thus have
\begin{eqnarray}
{\bf A}^{Qn}&=&
{\cal{R}}_{0\underline{n},+}{\bf A}^L{\cal{R}}^{-1}_{0\underline{n},+},
\nonumber\\
\label{I3plusconvention}
{\bf B}^{Qn}&=&
{\cal{R}}_{0\underline{n},+}{\bf B}^L{\cal{R}}^{-1}_{0\underline{n},+}.
\end{eqnarray}
The choice between the sets of ${\cal{R}}_{0n,+}$ and ${\cal{R}}_{0n,+}^{\dagger}$-induced transformations
is arbitrary, but once it is done, it has to be strictly adhered to.
The situation is analogous to a choice between
$(p_1,-p_3,+p_2)$,
obtained from $(p_1,p_2,p_3)$ through a rotation by $\pi/2$ 
around the first axis,
and
$(p_1,+p_3,-p_2)$,
obtained through a similar rotation by $-\pi/2$.
Each one of these two representations 
(or, in this case, also the original one, i.e. $(p_1,p_2,p_3)$) may be used.
The condition that the additivity of the 
momenta of different particles be properly
taken care of requires, however, that only one such form of description 
is universally  chosen for {\it all} particles.
For example, for the first convention,
the total momentum of a system of two
particles is properly calculated via the addition of the
corresponding representatives:
\begin{equation}
(p^{(1)}_1,-p^{(1)}_3,+p^{(1)}_2)+(p^{(2)}_1,-p^{(2)}_3,+p^{(2)}_2)=
(p^{tot}_1,-p^{tot}_3,+p^{tot}_2).
\end{equation}

\subsection{Quarks}
For the convention of Eq. (\ref{I3plusconvention}) and with the explicit forms of
${\bf A}^{Qn}$ and ${\bf B}^{Qn}$ given in
Eqs (\ref{AgenerSU4R}),
the matrices of the canonical momenta and positions 
for quarks of isospin $I_3=+1/2$
are, by analogy:
\begin{eqnarray}
P^Q\equiv 
\left[
\begin{array}{ccc}
{{\bf p}}^{Q1}\\
{\bf p}^{Q2}\\
{\bf p}^{Q3}
\end{array}
\right]
&=&
\left[
\begin{array}{ccc}
p^1_{1}&-x^1_{3}&+x^1_{2}\\
+x^2_{3}&p^2_{2}&-x^2_{1}\\
-x^3_{2}&+x^3_{1}&p^3_{3}
\end{array}
\right],\nonumber\\
&&\nonumber\\
\label{generSU4}
X^Q\equiv 
\left[
\begin{array}{ccc}
{{\bf x}}^{Q1}\\
{\bf x}^{Q2}\\
{\bf x}^{Q3}
\end{array}
\right]
&=&
\left[
\begin{array}{ccc}
x^1_{1}&-p^1_{3}&+p^1_{2}\\
+p^2_{3}&x^2_{2}&-p^2_{1}\\
-p^3_{2}&+p^3_{1}&x^3_{3}
\end{array}
\right],
\end{eqnarray}
where we have allowed that
the expressions on the right, containing the 
{\it physical} momenta and positions 
of quarks of different colours,  in general may depend on 
quark colour (hence superscript $n$ for row $n$). \\

 Transition to the sector of $I_3=-1/2$ quarks is obtained from 
 Eq. (\ref{I3plusconvention})
 by the action of ${\cal{R}}_{0\underline{n},+}$-induced 
 transformation upon $({\bf A}^{L'},{\bf B}^{L'})=({\bf A},-{\bf B})$, 
 corresponding to lepton $L'$ of isospin $I_3=-1/2$:
 \begin{eqnarray}
{\bf A}^{Q'n}&=&{\cal{R}}_{0\underline{n},+}
{\bf A}^{L'}{\cal{R}}^{-1}_{0\underline{n},+}
={\cal{R}}_{0\underline{n},+}
{\bf A}{\cal{R}}^{-1}_{0\underline{n},+},
\nonumber\\
\label{I3minusconvention}
{\bf B}^{Q'n}&=&{\cal{R}}_{0\underline{n},+}
{\bf B}^{L'}{\cal{R}}^{-1}_{0\underline{n},+}
={\cal{R}}_{0\underline{n},+}
(-{\bf B}){\cal{R}}^{-1}_{0\underline{n},+}.
\end{eqnarray}
 Thus,
\begin{eqnarray}
{A}^{Q'}
&=&
\left[
\begin{array}{ccc}
A_1&-B_3&+B_2\\
+B_3&A_2&-B_1\\
-B_2&+B_1&A_3
\end{array}
\right]=A^Q,\nonumber\\
&&\nonumber\\
{B}^{Q'}
&=&
-\left[
\begin{array}{ccc}
B_1&-A_3&+A_2\\
+A_3&B_2&-A_1\\
-A_2&+A_1&B_3
\end{array}
\right]=-B^Q.
\end{eqnarray}
The corresponding matrices of the canonical momenta 
and positions for quarks of isospin $I_3=-1/2$
are
\begin{eqnarray}
{P}^{Q'}
&=&
\left[
\begin{array}{ccc}
p^{1}_{1}&-x^{1}_{3}&+x^{1}_{2}\\
+x^{2}_{3}&p^{2}_{2}&-x^{2}_{1}\\
-x^{3}_{2}&+x^{3}_{1}&p^{3}_{3}
\end{array}
\right]=P^Q,\nonumber\\
&&\nonumber\\
\label{generSU4prime}
X^{Q'}
&=&
-\left[
\begin{array}{ccc}
x^{1}_{1}&-p^{1}_{3}&+p^{1}_{2}\\
+p^{2}_{3}&x^{2}_{2}&-p^{2}_{1}\\
-p^{3}_{2}&+p^{3}_{1}&x^{3}_{3}
\end{array}
\right]=-X^Q,
\end{eqnarray}
where, as before, we have allowed that
the expressions in the middle, containing the 
{\it physical} momenta and positions 
of quarks of different colours,  in general may depend on 
quark colour. On the other hand, 
for better transparency of the comparison of the $I_3=-1/2$ and 
$I_3=+1/2$ sectors,
we have suppressed the superscript `prime' on physical variables.

Applying
the operation of phase-space reflection defined in
Eq. (\ref{phspreflection})
 to the set $[(A^{Q},B^{Q}),$ $(A^{Q'},B^{Q'})]$ 
(corresponding to [$I_3=+1/2$, $I_3=-1/2$]) 
 yields
\begin{eqnarray}
I_1(A^{Q},B^{Q})I_1^{-1}&=&(A^{Q'(\dagger)},B^{Q'(\dagger)}),\nonumber\\
I_1(A^{Q'},B^{Q'})I_1^{-1}&=&(A^{Q(\dagger)},B^{Q(\dagger)}),
\end{eqnarray}
i.e. one obtains representatives for 
quarks with interchanged eigenvalues of isospin 
($\pm 1/2 \to \mp 1/2$), but in the other representation.
This is so because
$I_1{\cal{R}}_{0n,+}I_1^{-1}={\cal{R}}_{0n,+}^{(\dagger)}$, and is related to a
change from a left- to a right-handed labelling of the system of coordinates.

To summarize:
 just as leptons of isospin $I_3=\pm1/2$ correspond to
 $({\bf A}^L,\pm {\bf B}^L) $
 and $({\bf p}^L,\pm {\bf x}^L)$ , so do quarks of isospin $I_3=\pm 1/2$
 correspond to $({A}^Q,\pm {B}^Q)$ and $({P}^Q,\pm X^Q)$.
 Note that the forms of the canonical momenta ($P^Q)$ are identical for
 both isospins.

\subsection{Antiquarks}
We now recall that the transition to antiparticles   
 is obtained via complex conjugation of Eq. (\ref{ChconjClAlg}). 
 In a lepton sector, this leads to
 $({\bf {A}^L}, {\bf {B}}^L,i)=({\bf A},{\bf B},i)\to 
 ({\bf {A}}^{\overline{L}}, {\bf {B}}^{\overline{L}},i^*)= 
 ({\bf A}, -{\bf B},-i)$
 and
 $({\bf {p}^L}, {\bf {x}}^L)=({\bf p},{\bf x})\to 
 ({\bf {p}}^{\overline{L}}, {\bf {x}}^{\overline{L}})= 
 ({\bf p}, -{\bf x})$.
In the quark sector,  
this leads to the following matrices for 
the counterparts of canonical momenta and positions of antiquarks:\\

1) for the $\bar{I}_3=-1/2$ antiparticles of the $I_3=+1/2$ quarks 
 \begin{eqnarray}
A^{\overline{Q}}
&=&
\left[
\begin{array}{ccc}
A_1&+B_3&-B_2\\
-B_3&A_2&+B_1\\
+B_2&-B_1&A_3
\end{array}
\right],\nonumber\\
&&\nonumber\\
\label{AbargenerSU4R}
B^{\overline{Q}}
&=&
\left[
\begin{array}{ccc}
-B_1&-A_3&+A_2\\
+A_3&-B_2&-A_1\\
-A_2&+A_1&-B_3
\end{array}
\right],
\end{eqnarray}
and
 \begin{eqnarray}
P^{\overline{Q}}
&=&
\left[
\begin{array}{ccc}
{p}^1_{1}&+{x}^1_{3}&-{x}^1_{2}\\
-{x}^2_{3}&{p}^2_{2}&+{x}^2_{1}\\
+{x}^3_{2}&-{x}^3_{1}&{p}^3_{3}
\end{array}
\right],\nonumber\\
&&\nonumber\\
\label{generSU4anti}
X^{\overline{Q}}
&=&
\left[
\begin{array}{ccc}
-{x}^1_{1}&-{p}^1_{3}&+{p}^1_{2}\\
+{p}^2_{3}&-{x}^2_{2}&-{p}^2_{1}\\
-{p}^3_{2}&+{p}^3_{1}&-{x}^3_{3}
\end{array}
\right].
\end{eqnarray}

2) for the $\bar{I}_3=+1/2$ antiparticles of the $I_3=-1/2$ quarks:\\
 \begin{eqnarray}
A^{\overline{Q}'}
&=&
\left[
\begin{array}{ccc}
A_1&+B_3&-B_2\\
-B_3&A_2&+B_1\\
+B_2&-B_1&A_3
\end{array}
\right]=A^{\overline{Q}},\nonumber\\
&&\nonumber\\
\label{AbargenerSU4Ranti}
B^{\overline{Q}'}
&=&
-\left[
\begin{array}{ccc}
-B_1&-A_3&+A_2\\
+A_3&-B_2&-A_1\\
-A_2&+A_1&-B_3
\end{array}
\right]=-B^{\overline{Q}},
\end{eqnarray}
and
\begin{eqnarray}
P^{\overline{Q}'}
&=&
\left[
\begin{array}{ccc}
{p}^1_{1}&+{x}^1_{3}&-{x}^1_{2}\\
-{x}^2_{3}&{p}^2_{2}&+{x}^2_{1}\\
+{x}^3_{2}&-{x}^3_{1}&{p}^3_{3}
\end{array}
\right]=P^{\overline{Q}},\nonumber\\
&&\nonumber\\
\label{generSU4primeanti}
X^{\overline{Q}'}
&=&
-\left[
\begin{array}{ccc}
-{x}^1_{1}&-{p}^1_{3}&+{p}^1_{2}\\
+{p}^2_{3}&-{x}^2_{2}&-{p}^2_{1}\\
-{p}^3_{2}&+{p}^3_{1}&-{x}^3_{3}
\end{array}
\right]=-X^{\overline{Q}}.
\end{eqnarray}
The pattern of {\it relative} signs between all of Eqs 
(\ref{generSU4}, \ref{generSU4prime}, \ref{generSU4anti}, 
\ref{generSU4primeanti}) is closely connected to the structure of
the eigenvalues of
$I_3$ and $Y$.
Had we started from the other representation of the algebraic counterparts
of the $I_3=+1/2$ quarks, i.e. from $(A^{Q(\dagger)},B^{Q(\dagger)})$, the
relevant phase space
representations for all other quarks and antiquarks would have changed
accordingly. In particular, 
this would amount to a change of sign in front of all physical position
coordinates appearing in $P^Q$, $P^{Q'}$, $P^{\bar{Q}}$, and $P^{\bar{Q'}}$.
The relative signs between the components in a given direction in phase space would
have stayed unchanged, however.
Thus, e.g. if $x^n_k$ enters into canonical momenta of a quark of colour $n \ne
k$ with a given (positive or negative) sign, then $x^m_k$ 
(for $n \ne m\ne k$) enters
with the opposite sign.
Furthermore,
the relative connection between quarks and antiquarks is
independent of whether we start from $(A^Q,B^Q)$ or from 
$(A^{Q(\dagger)},B^{Q(\dagger)})$.
Just as for leptons and antileptons,
the quarks and antiquarks  are connected by
a reflection in (physical) position space
(see Eqs (\ref{generSU4},\ref{generSU4anti}) 
or Eqs (\ref{generSU4prime},\ref{generSU4primeanti})), which may be
symbolically written  as:
\begin{eqnarray}
P^{\overline{Q}}({p}^n_k,{x}^m_l)&=&P^{{Q}}({p}^n_k,-{x}^m_l),\nonumber\\
X^{\overline{Q}}({p}^n_k,{x}^m_l)&=&X^{{Q}}({p}^n_k,-{x}^m_l),
\end{eqnarray} 
accompanied by the change of sign of the free-standing $i$.
This, together with Eqs (\ref{antileptonconv},\ref{antipx1}) for leptons, is the phase-space counterpart of the standard interpretation of 
 antiparticles as `particles moving backward in time'. Yet,
 in the phase-space-based interpretation the concept of
  `time' is not introduced in any explicit way.
 This seems to fit well into the philosophy of timeless 
 quantum gravity.

\section{Discussion and outlook} 
  In the classical limit, when one goes
   with the Planck constant to zero, 
   the $i$ ceases to contribute on the r.h.s. of 
   position-momentum commutation relations.   
     With
     the quantum $i$ absent, 
     the difference between particles and antiparticles (of a given isospin) -
     when interpreted
   in terms of classical phase-space concepts -  
     reduces to the reflection of position space alone. 
       Note that this might have been expected on the basis of the appearance -
     already in the case of the {\it classical} 3D harmonic oscillator (i.e. no
     $i$) -
     of the $U(1) \otimes SU(3)$ symmetry group.
     There is nothing wrong with the existence of two different classical
     interpretations of the connection between particles and antiparticles: the
     St\"uckelberg-Feynman interpretation and 
     the phase-space interpretation herein proposed 
     simply constitute different faces of the same coin.
     The advantage of the phase-space picture is that it permits a {\it
     timeless}
     classical interpretation of the connection between particles and antiparticles,
     with timelessness being a feature deemed welcome
     for the development of
    quantum gravity
     \cite{Rovelli1991},\cite{Kiefer}.
 
A very interesting feature of Eqs (\ref{generSU4},\ref{generSU4prime}) and their
charge-conjugate versions of Eqs (\ref{generSU4anti},\ref{generSU4primeanti}) is the pattern of signs in front of the physical
components of phase-space variables (e.g. $-x^1_{3}$ and $+x^1_{2}$ in Eq. 
(\ref{generSU4})).
The negative signs cannot be all simultaneously changed into 
positive ones by a
redefinition of the type: 
\begin{equation}
\label{redef}
(x^1_{3},x^1_{2}) \to ({x'}^1_{3},{x'}^1_{2})\equiv (-x^1_{3},x^1_{2}),
\end{equation}
 because
 the preservation of the physical content
of Eq. (\ref{generSU4}) requires that
such a redefinition be applied to all relevant  
position and momentum components simultaneously
(i.e., in the case of
the redefinition of Eq. (\ref{redef}),
to $x^n_{3}$, $x^n_{2}$ with any $n$, 
and to the respective components of momenta).
This pattern of signs follows from our physical assumptions 
(recall also the discussion 
just before Section \ref{sectionRn+}).

Keeping in mind the pattern of signs in Eqs
(\ref{generSU4},\ref{generSU4prime},\ref{generSU4anti},\ref{generSU4primeanti}), 
we now come to the issue of the additivity of quark momenta.
As already stressed in the introduction, the
 physical momentum of a 
system of ordinary classical particles 
 is obtained by {\it adding}
the physical momenta of its components, independently of where
(in the configuration space) the
particles are located. 
This additivity carries over to the standard quantum formalism.
In our approach there exist two seemingly `natural', but different, ways 
in which the idea of the additivity
of physical momenta of ordinary particles may be generalized to the quark
sector. The two generalizations exist because 
the concepts of physical
and canonical momenta coincide here for ordinary particles, 
but are not equivalent for quarks.
One may therefore consider either the additivity of quark physical momenta,
or the additivity of quark canonical momenta. 
The latter option is possible because 
- despite the theoretical edifices built -
the additivity of physical
momenta of {(standardly defined)} {individual} quarks 
has not been really tested,
 as these momenta are
{\it never measured}.
Indeed, we always measure the momenta of quark conglomerates (i.e. hadrons).
In standard approaches we then assume that hadrons are built of quarks
possessing properties of ordinary particles, and in particular, satisfying the
standard way of ensuring the additivity of quark momenta.
Now, the whole argument of bringing more symmetry between physical
position and momentum variables suggests that it is the additivity of 
{\it canonical} momenta which
is more natural within our scheme.
Consequently, let us accept that 
{\it the additivity of the (operators of)
 physical momenta of 
{ordinary} particles (and antiparticles)
is a special case of the general principle of additivity of (the operators of) 
canonical momenta}. Then, 
one can apply the concept of additivity of canonical momenta to the quark sector. 

In order to gain some quasi-classical understanding of the situation, consider
a quark-antiquark 
$q_{\underline{n}}\bar{q}_{\underline{n}}$ system for a fixed $n$ 
(i.e. no superpositions of $q\bar{q}$ pairs 
of different colour) and with a well-defined canonical momentum. 
Its description should then
involve, in particular,
{ordinary} addition of
quark and antiquark physical momenta in the $n$-th direction.  
Due to the opposite signs with which quark and antiquark positions (for any
combination of $I_3$ eigenvalues)
enter into the expressions for the canonical momenta ${\bf p}^{Qn}$, 
${\bf p}^{\bar{Q}n}$,
the addition of the appropriate components of the latter
leads to the {\it subtraction} of the corresponding
physical position coordinates in directions perpendicular 
to the physical momentum.
Hence, while individual quarks possess translationally noninvariant
canonical momenta, 
 the translational invariance of their sum, relevant for
the composite $q\bar{q}$ system, is ensured. 
This has some similarity to the idea of a translationally invariant
string connecting
quark and antiquark.

Another way of forming translationally invariant  
expressions from quark canonical
momenta is through
the addition of canonical momenta of three quarks of different colours. 
In this case - again thanks to the different signs with 
which the physical positions enter 
into the expressions for the canonical momenta - 
all three quarks will conspire together to form translationally invariant
expressions, i.e. $x^2_3-x^1_3$, $x^1_2-x^3_2$, $x^3_1-x^2_1$ from 
Eq. (\ref{generSU4}). 
(The idea of quark conspiracy requires also that phase-space variables
corresponding to quarks of different colours, when appropriately grouped,
form ordinary vectors together.)
On the other hand, a pair of quarks is not sufficient to
form translationally invariant expressions.
Clearly, all this looks just like 
 the emergence of mesons and baryons as composite systems built
 from unobservable 
 quarks.

Note that a classical picture requires talking about
well-defined canonical positions in addition to well-defined canonical momenta.
Here, however,
in both meson and baryon cases we are not dealing with 
well-defined canonical positions
 as yet. Thus,
in fact, we do not have
 strings `stretching between' quarks:  our quarks are still
not fully localized in the ordinary 3D configuration
space.
The picture is still quantum.

In order to see if `real' strings can be obtained, one would need
a better understanding of the transition from the quantum to the classical 
description of Nature. This is the goal of all emergent-space programs including
our `emergent phase space' proposal, and 
falls obviously far beyond the scope of this paper. 
The quasi-classical limit of the quantum picture to which we are led is certainly weird.
Yet, it follows in a logical way from fundamental assumptions
which, in my opinion, look very natural.
Consequently, I am inclined to believe that it captures an important aspect of 
physical reality.

The above discussion is intended to show that our proposal for the nature
of quarks has a built-in capacity to explain the phenomenon of quark 
confinement within
a string-{\it like} description.
Such a description does not have to be in conflict with the
present field-theoretical QCD approach, which describes
what happens at large momenta transfers, not at large position differences,
and does it within a specific theoretical description framework, 
 built upon the {\it
 background} of ordinary 3D space (or spacetime).
In our approach,
on the other hand, this classical background 
(together with the relevant gauge structure)
is expected to
 emerge only later, and it is only then that more precise
 connections to the QCD description may be sought.

Addivity of ordinary physical momenta may be viewed 
as resulting 
from a limiting case of the
additivity of angular momenta,
when the point with respect to which angular momenta are
evaluated is shifted to spatial infinity.
 Thus, additivity of physical momenta 
 may be traced back
to the quantum-level additivity of the spin operators of component particles. 
Analogously, additivity of
the canonical momenta in the
quark sector should also have a corresponding operator
counterpart at the discrete quantum level. It is
this counterpart that would presumably play an important r\^ole in the
construction of properly behaving protohadronic composite systems.

\section{Summary}
In this paper we have proposed a phase-space interpretation of the connections
between leptons, quarks, and their antiparticles. The interpretation is timeless
$-$ which might be relevant for quantum gravity $-$ and amounts
to different relations between physical and canonical phase-space variables.
In particular, some components of the canonical momenta of quarks are
identified with their physical positions, thus lacking invariance under
translations.
We suggested
that the principle of the additivity of physical momenta for leptons 
(and other ordinary particles)
is a special case of the additivity of canonical momenta in general.
This generalization was then shown to lead to the emergence 
of translationally invariant expressions for the $q\bar{q}$ and $qqq$ systems,
a mechanism conjectured to be intimately related to quark confinement
and the existence of mesons and baryons.\\

I would like to thank Enrico Prati 
for bringing the FQXi contest
on the Nature of Time to my attention.

\end{document}